\RequirePackage{lineno}
\documentclass[twocolumn,aps,prc,showpacs,superscriptaddress,preprintnumbers,floatfix,nofootinbib]{revtex4}
\usepackage{epsfig,graphics}
\usepackage{graphicx}
\usepackage{dcolumn}
\usepackage{bm}
\usepackage{amsmath}
\usepackage[usenames]{color}
\usepackage{ulem} 
\usepackage{url} 

\begin{document}

\title{Medium modifications of jet shapes in Pb+Pb collisions at $\sqrt{s_{_{\rm NN}}}$ = 2.76 TeV within a multiphase transport model}

\author{Guo-Liang Ma}
\affiliation{Shanghai Institute of Applied Physics, Chinese
Academy of Sciences, P.O. Box 800-204, Shanghai 201800, China}


\begin{abstract}
Within a multiphase transport model, medium modifications of differential jet shapes are investigated in Pb+Pb collisions at $\sqrt{s_{_{\rm NN}}}$ = 2.76 TeV. The differential jet shapes are significantly modified by the strong interactions between jets and a partonic medium in Pb+Pb collisions relative to that in p+p collisions. The modifications are slightly weakened by the hadronization of coalescence, but strengthened by resonance decays in hadronic rescatterings. Subleading jets display larger medium modifications than leading jets, especially in central Pb+Pb collisions with large dijet asymmetries. These behaviors of medium modifications of differential jet shapes reflect a dynamical evolution of redistribution of jet energy inside a quenched jet cone in high-energy heavy-ion collisions. 
\end{abstract}


\maketitle

\section{Introduction}
\label{sec:intro}
One important discovery at the Relativistic Heavy Ion Collider (RHIC) and the Large Hadron Collider (LHC) is that jets, produced by initial QCD hard scatterings with high transverse momentum $p_{T}$, are strongly quenched because they interact with the formed partonic medium in high-energy heavy-ion collisions~\cite{Wang:1991xy}. It has been proved by many observations such as the strong suppressions of nuclear modification factors $R_{AA}$ and $I_{AA}$ from single-hadron~\cite{Aamodt:2010jd} and dihadron~\cite{Aamodt:2011vg} measurements, respectively. The recent measurements based on reconstructed jets at the LHC, such as dijet $p_{T}$ asymmetry~\cite{Aad:2010bu, Chatrchyan:2011sx}, $\gamma$-jet $p_{T}$ imbalance~\cite{Chatrchyan:2012gt}, and jet fragmentation function~\cite{Chatrchyan:2012gw, ATLAS:2012ina, CMS:2012wxa,Chatrchyan:2013kwa}, provide us good chances to comprehensively investigate the jet energy loss mechanisms. One impressive measurement is medium modification of jet shape, which measures the medium modifications of the probability distribution of transverse energy carried by associated particles inside a quenched jet cone in Pb+Pb collisions. The experimental results show no modification at a small radius but large enhancement at a large radius in central Pb+Pb collisions, relative to that in p+p collisions~\cite{CMS:2012wxa,Chatrchyan:2013kwa}. To my knowledge, few theoretical studies have been done to interpret it so far. Vitev $et~al.$ predicted the medium-induced enhancement of jet shape can reach $\sim$ 1.75 in the periphery of the cone for central Pb+Pb collisions at $\sqrt{s_{_{\rm NN}}}$ = 5.5 TeV using their framework of perturbative QCD and jet radiative energy loss~\cite{Vitev:2008rz}, and proposed reconstructed jets as sensitive tomographic probes of the quark-gluon plasma (QGP)~\cite{Vitev:2009rd}. Wang and Zhu~\cite{Wang:2013cia} recently found that the induced gluon radiations play a more significant role to broaden the differential $\gamma$-jet shape than elastic parton scatterings within a linearized Boltzmann transport model without including hadronization or hadronic phase evolution. In this paper, the medium modifications of differential jet shapes are investigated in Pb+Pb collisions at $\sqrt{s_{_{\rm NN}}}$ = 2.76 TeV within a multiphase transport (AMPT) model, which includes both dynamical evolutions of partonic and hadronic phases. The differential jet shapes are largely modified by the strong interactions between jets and a partonic medium. The additional effects of hadronization and hadronic rescatterings are also discussed. Subleading jets display more significant medium modifications than leading jets, especially in central Pb+Pb collisions with large dijet asymmetries. These medium modifications of differential jet shapes reveal how the jet energy is redistributed inside a quenched jet cone through the interactions between jet and the hot medium in high-energy heavy-ion collisions. 

\section{The AMPT Model}
\label{sec:model}
The AMPT model with string melting scenario is utilized in this work~\cite{Lin:2004en}. It consists of four main stages of high-energy heavy-ion collisions: the initial condition, parton cascade, hadronization, and hadronic rescatterings. In order to study the jet energy loss behaviors, a dijet of $p_{T}\sim$ 90 GeV/$c$ is triggered in the initial condition based on the heavy-ion jet interaction generator (HIJING) model~\cite{Wang:1991hta,Gyulassy:1994ew}. The high-$p_{T}$ primary partons pullulate into jet showers full of lower virtuality partons through initial- and final- state QCD radiations. The jet parton showers are converted into clusters of on-shell quarks and anti-quarks through the string melting mechanism of the AMPT model. After the melting process, both a quark and anti-quark plasma and jet quark showers are built up. Next, Zhang's Parton Cascade (ZPC) model~\cite{Zhang:1997ej} automatically simulates all possible elastic partonic interactions among medium partons and jet shower partons, but without including inelastic parton interactions or further radiations at present. When the partons freeze out, they are recombined into medium hadrons or jet shower hadrons via a simple coalescence model which combines two nearest quarks into a meson and three nearest quarks into a baryon. The final-state hadronic interactions including baryon-baryon, baryon-meson, and meson-meson elastic and inelastic scatterings and resonance decays such as $\rho$ mesons, $\Delta$ and $N^{*}$ baryons, can be described by a relativistic hadronic transport (ART) model~\cite{Li:1995pra}. Recently, the AMPT model with a partonic interaction cross section of 1.5 mb has successfully given many qualitative descriptions of the experimental results about pseudorapidity and $p_{T}$ distributions~\cite{Xu:2011fi}, harmonic flows~\cite{Xu:2011fe, Xu:2011jm}, and reconstructed jet observables, such as $\gamma$-jet $p_{T}$ imbalance~\cite{Ma:2013bia}, dijet $p_{T}$ asymmetry~\cite{Ma:2013pha}, and jet fragmentation function~\cite{Ma:2013gga} in Pb+Pb collisions at the LHC energies. Consistently with the previous studies, two sets of partonic interaction cross sections, 1.5 mb and 0 mb, are respectively chosen to simulate Pb+Pb and p+p collisions at $\sqrt{s_{_{\rm NN}}}$ = 2.76 TeV in this work.
\section{Jet Reconstruction}
\label{sec:jetrec}
To reconstruct jets, the kinematic cuts are chosen to be the same as in the CMS experiment~\cite{CMS:2012wxa,Chatrchyan:2013kwa}. The transverse momentum of a jet is required to be larger than 100 GeV/$c$ within a pseudorapidity $\eta$ range of $0. 3 < |\eta|<2$ for this analysis. Jets within $|\eta|<0.3$ are excluded in order to avoid the overlap between the signal jet region and the jet background estimation region. The anti-$k_{t}$ algorithm from the standard Fastjet package~\cite{Cacciari:2011ma} is used to reconstruct full jets in which the jet cone sizes $R$ are set to be 0.3. The differential jet shape $\rho (r)$ is defined as the fraction of the transverse momentum carried by particles ($p_{T}>$ 1 GeV/$c$) associated with the jet, which are contained inside an [$\eta$, $\phi$ (azimuthal angle)] annulus of inner and outer radii of $r \pm \delta r/2$ around the jet axis, where $\delta r$ is chosen to be 0.05 and $0 \le r \le R$. $\rho (r)$ satisfies the normalization condition $\int_{0}^{R}\rho (r) dr$ = 1. The $\eta$-reflection method as the CMS experiment did, i.e. selecting the particles that lie in a background jet cone obtained by reflecting the original jet cone around $\eta$=0 while keeping the same $\phi$ coordinate, is used to estimate the background, which is subtracted from the reconstructed differential jet shape.

\section{Results and Discussions}
\label{sec:results}

\begin{figure}
\includegraphics[scale=0.45]{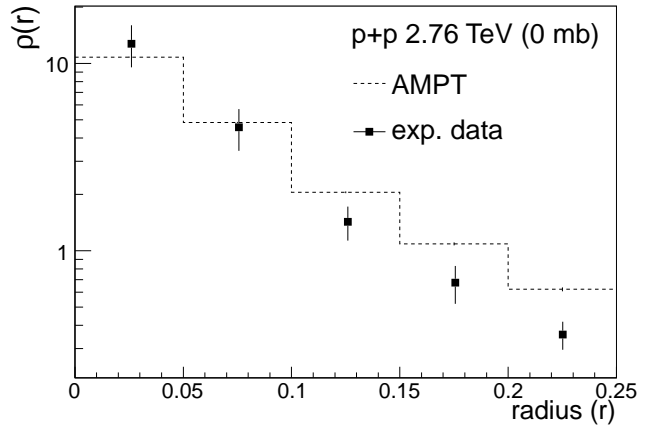}
\caption{The differential jet shapes $\rho(r )$ in p+p collisions at $\sqrt{s_{_{\rm NN}}}$ = 2.76 TeV, where the histogram represents the AMPT result with hadronic interactions only, and the squares represent the data from the CMS experiment~\cite{CMS:2012wxa,Chatrchyan:2013kwa}. }
 \label{fig-pp}
\end{figure}

Figure~\ref{fig-pp} shows the comparison of the differential jet shapes $\rho(r )$ between the AMPT result with hadronic interactions only (with a partonic interaction cross section of 0 mb) and experimental data for p+p collisions at $\sqrt{s_{_{\rm NN}}}$ = 2.76 TeV. The AMPT result basically can describe p+p data, which provides a qualified baseline for the following calculations in Pb+Pb collisions.
 
\begin{figure}
\includegraphics[scale=0.45]{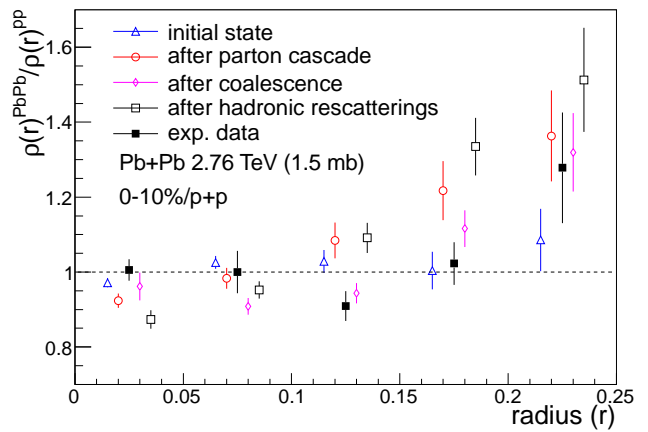}
\caption{(Color online) The differential jet shape ratios of most central Pb+Pb collisions (0-10\%) to p+p collisions, where the open symbols represent the ratios at different evolution stages from the AMPT simulations with both partonic and hadronic interactions, and the solid squares represent the data from the CMS experiment~\cite{CMS:2012wxa,Chatrchyan:2013kwa}. Some points are slightly shifted along the $x$ axis for better representation.
}
 \label{fig-evo}
\end{figure}

As heavy-ion collisions are dynamical evolutions which involve many important evolution stages, it is necessary to investigate how the distribution of jet transverse energy develops. Figure~\ref{fig-evo} shows the differential jet shape ratios $\rho(r)^{PbPb}/\rho(r)^{pp}$ of most central Pb+Pb collisions (0-10\%) to p+p collisions at different evolution stages from the AMPT simulations with both partonic (with a partonic interaction cross section of 1.5 mb) and hadronic interactions, in comparison with the experimental data. The initial ratios are consistent with unity, which indicates no jet shape modification in the initial state of Pb+Pb collisions compared to the jet shape in p+p collisions. However, the differential jet shape is significantly modified after the parton cascade process in which jet shower partons interact with medium partons frequently. The emergent modifications show a suppression at a small radius and an enhancement at a large radius, which implies the jet energy is redistributed towards a larger radius via the strong interactions between the jet and the partonic medium. It is consistent with Wang and Zhu's results with elastic scatterings only, however they found an even larger enhancement of differential jet shape can be produced if including inelastic scatterings~\cite{Wang:2013cia}. The coalescence mechanism can recombine the jet shower partons and medium partons into jet shower hadrons. This reduces the modifications of jet shape and gives qualitative features of the experimental data after the hadronization. My recent work suggested that there may exist a competition between coalescence and fragmentation for jet hadronization in high-energy heavy-ion collisions~\cite{Ma:2013gga}, but only the coalescence mechanism is considered here for simplicity. However, the modifications of jet shape are strengthened after hadronic rescatterings, because resonance decays make a smearing effect to push the jet energy outward further. Compared to experimental data, the final  AMPT result underestimates the data for the small radius range and overestimates the data for the large radius range in magnitude.

\begin{figure}
\includegraphics[scale=0.45]{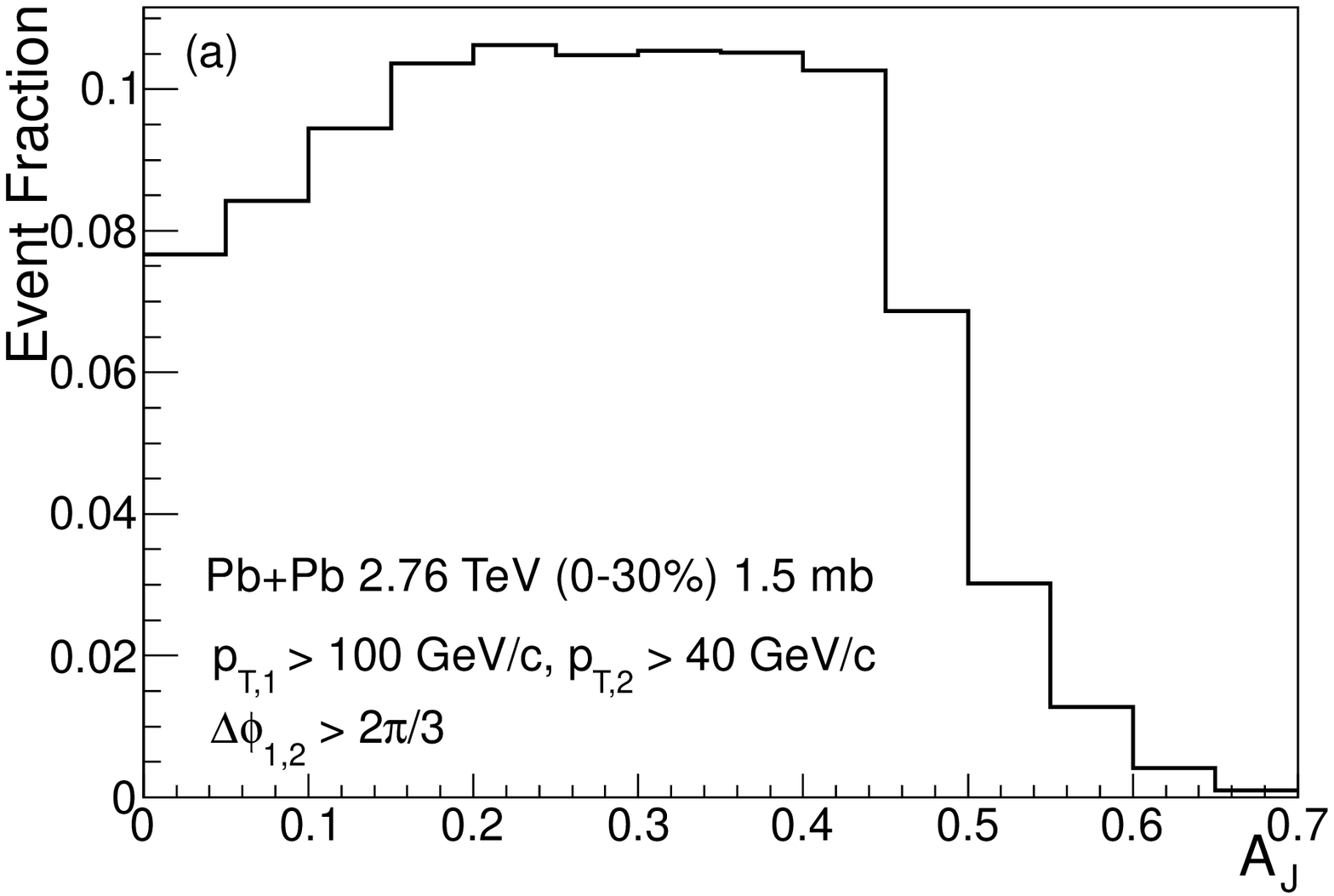}
\includegraphics[scale=0.45]{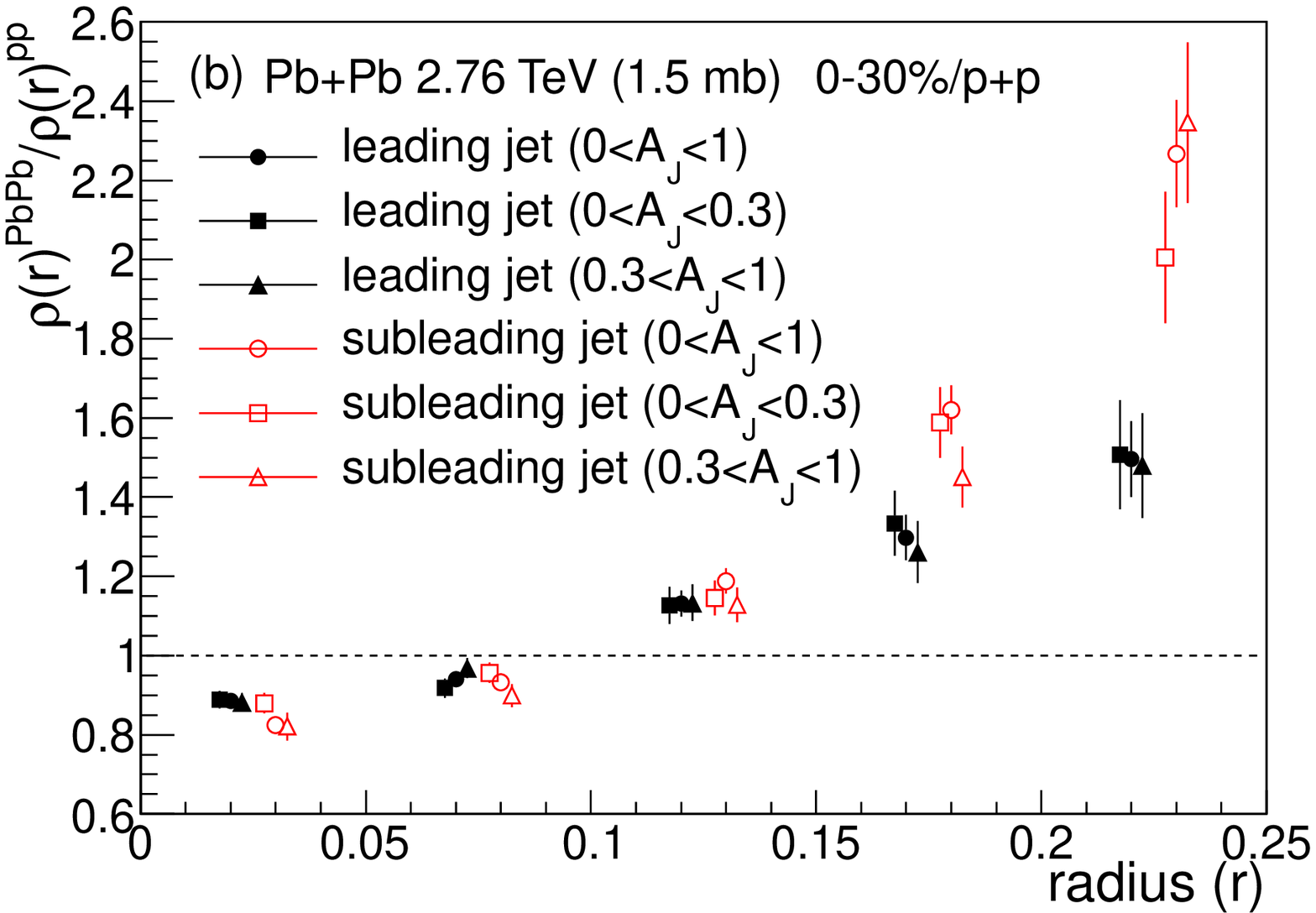}
\caption{(Color online) (a) The AMPT results on dijet asymmetry ratio $ A_{J}$ distribution for the centrality bin of 0-30\% in Pb+Pb collisions at $\sqrt{s_{_{\rm NN}}}$ = 2.76 TeV. (b) The differential leading and subleading jet shape ratios of the centrality bin of 0-30\% in Pb+Pb collisions to p+p collisions with different dijet asymmetry ratio $ A_{J}$ selections. Some points are slightly shifted along the $x$ axis for better representation.
}
\label{fig-aj}
\end{figure}

Recently, a large dijet transverse momentum asymmetry has been observed by the LHC experiments~\cite{Aad:2010bu,Chatrchyan:2011sx}, which have been well described by the AMPT model~\cite{Ma:2013pha}. Figure~\ref{fig-aj} (a) shows the AMPT result on a dijet asymmetry ratio $A_{J}$=($p_{T,1}-p_{T,2}$)/($p_{T,1}+p_{T,2}$)  distribution for leading jets of $p_{T,1} >$ 100 GeV/$c$ with subleading jets of $p_{T,2} >$ 40 GeV/$c$, and $\Delta\phi_{1,2} > 2\pi/3$ for the centrality bin of 0-30\% in Pb+Pb collisions at $\sqrt{s_{_{\rm NN}}}$ = 2.76 TeV, where the subscripts 1 and 2 refer to the leading jet and subleading jet respectively. It is clearly seen that the large dijet $p_{T}$ asymmetry between leading and subleading jets can be reproduced owing to jet partonic energy loss in a partonic medium in the central Pb+Pb collisions by the AMPT model. Therefore, it becomes possible to measure the medium modifications of differential jet shapes in more detail with the help of the dijet asymmetry ratio $A_{J}$.  Figure~\ref{fig-aj} (b) shows the differential leading and subleading jet shape ratios of the centrality bin of 0-30\% in Pb+Pb collisions to p+p collisions with different dijet asymmetry ratio $ A_{J}$ selections, where an arbitrary cut of $A_{J}$=0.3 is chosen to divide all dijet events into two parts.  Basically, the differential jet shapes for subleading jets are more strongly suppressed for a small radius and more highly enhanced for a large radius, compared to those for leading jets. It can be easily understood as the fact that subleading jets always lose more energy than leading jets because of their longer path lengths. With the further $A_{J}$ selections, the differential subleading jet shape shows a slightly stronger modification with the large $A_{J}$ selection (0.3 $<A_{J}<$ 1) than that with the small $A_{J}$ selection (0 $<A_{J}<$ 0.3), which indicates that the jet shape is more modified owing to the increasing of subleading jet energy loss.

\section{Summary}
\label{sec:summary}

In summary, the medium modifications of differential jet shapes are investigated in Pb+Pb collisions at $\sqrt{s_{_{\rm NN}}}$ = 2.76 TeV in the framework of a multiphase transport (AMPT) model. The significant modifications of differential jet shapes are produced by the strong interactions between jets and a partonic medium. The modifications are then reduced by the hadronization of coalescence, but strengthened by resonance decays in hadronic rescatterings. Subleading jets show larger medium modifications of differential jet shapes than leading jets, especially for large-$A_{J}$ events. These behaviors of modifications of differential jet shapes reflect a dynamical evolution of redistribution of jet energy inside a quenched jet cone in high-energy heavy-ion collisions. 

\section*{ACKNOWLEDGMENTS}

This work was supported by the Major State Basic Research Development Program in China under Contract No. 2014CB845404, the NSFC of China under Projects No. 11175232, No. 11035009, and No. 11375251, the Knowledge Innovation Program of CAS under Grant No. KJCX2-EW-N01, the Youth Innovation Promotion Association of CAS, the project sponsored by SRF for ROCS, SEM, CCNU-QLPL Innovation Fund under Grant No. QLPL2011P01, and the ``Shanghai Pujiang Program" under Grant No. 13PJ1410600.

\end{document}